\documentclass{eas}
\usepackage{graphicx}
\usepackage{amsmath} 
\usepackage{amssymb} 
\usepackage{mathrsfs}
\begin{document}

\title{The {\it Gaia} spectroscopic instrument (RVS): a technical challenge} 

\runningtitle{RVS: a technical challenge}
\author{Mark Cropper}\address{Mullard Space Science Laboratory, University College London, Holmbury St Mary, Dorking, Surrey RH5 6NT, United Kingdom}
\author{David Katz}\address{GEPI, Observatoire de Paris Meudon, 5 place Jules Janssen, 92190 Meudon, France}
%
%
\begin{abstract}
The lack of radial velocity data in the {\it Hipparcos} catalogue was considered a significant deficiency, so when {\it Gaia} was conceived, a spectrometer was a core constituent of its payload. The {\it Gaia} Radial Velocity Spectrometer faced a number of design challenges, in particular set by the need to balance kinematic and astrophysical capability. We present an overview of the evolution of the instrument to its present form, identifying the competing technical, performance and programmatic factors which have shaped it.
\end{abstract}
\maketitle


\section{Introduction}
By measuring the positions of the stars, and as these change over time, by calculating their transverse velocities, their energies and momenta can be calculated, and the mass distribution within the Galaxy and its assembly history may be understood. The distance of the star can also be determined, using the same measurements, by triangulation from measurements six months apart (parallax). But to determine fully the energies and momenta, one remaining velocity is required to complete the full set of three positions and three velocities: that in the line of sight. Although some progress can be made from the astrometric measurements alone (Dravins {\it et al.} 1999), in practice this is best measured by the Doppler shift, and it requires a different instrument -- a spectrometer.

The absence of a radial velocity spectrometer from the premier astrometry mission of its time, {\it Hipparcos}, was a matter of concern to the scientific community from the beginning of that mission, in 1980. However, while large, the total number of stars in the {\it Hipparcos} Input Catalog was not too large for it to be contemplated that these measurements could be made using ground-based spectrometers. Proposals were submitted in 1980, 1981 and 1982 in France, and then in 1987 in the United Kingdom, to build dedicated instruments in both Northern and Southern hemispheres (Turon, private communication; Perryman, 2009). These projects were of moderate cost compared to that of {\it Hipparcos}, but competed for other resources and none of the bids were successful, despite being strongly supported within the {\it Hipparcos} scientific community. In his summary remarks at the 1988 Sitges meeting, Blaauw again stressed the urgency of attending to this deficiency. 

The only way forward was with existing instrumentation, and large surveys were initiated at ESO in the South, and at Observatoire  de Haute-Provence in the North. Observations over more than 400 nights were carried out at ESO, and in total data were obtained for more than a third of the {\it Hipparcos} catalog. Unfortunately there is no complete publication of this large resource, although Nordstr\"{o}m {\it et al.} (2004) and Famaey {\it et al.} (2005) made use of part of it (the former having started with a separate Geneva-Copenhagen programme of F-star observations) (Turon, private communication). The {\it Hipparcos}  Catalog was published in 1997 without associated radial velocities. Perryman (2003, 2009) provides a more comprehensive overview of the developments at those times.

Hence, when the concept for a successor to {\it Hipparcos} was initiated, the inclusion of a spectrometer was high on the priority list. In their 1995 paper, Favata \& Perryman described slitless scanning spectrometers for both ground and space to support the {\it Gaia} mission. By 1997, aware of the huge number of spectra that would be required, and the practical difficulties faced by the earlier efforts described above, they had developed their concept to being a space-based instrument only, to be called ARVI (Absolute Radial Velocities Instrument), incorporated into the {\it Gaia} payload (Favata \& Perryman 1997). These two papers identified the main issues to be considered for realising this instrumentation. They adopted a scheme with a dedicated wide-field telescope of diameter 0.9m to feed the spectrometer, which was to operate in a narrow spectral range around the MgII complex, and they identified the importance of readout noise in the CCDs as a critical issue. Their performance predictions are close to those expected for the final {\it Gaia} spectrometer.

\begin{figure}
\center{
\includegraphics[height=5.5cm]{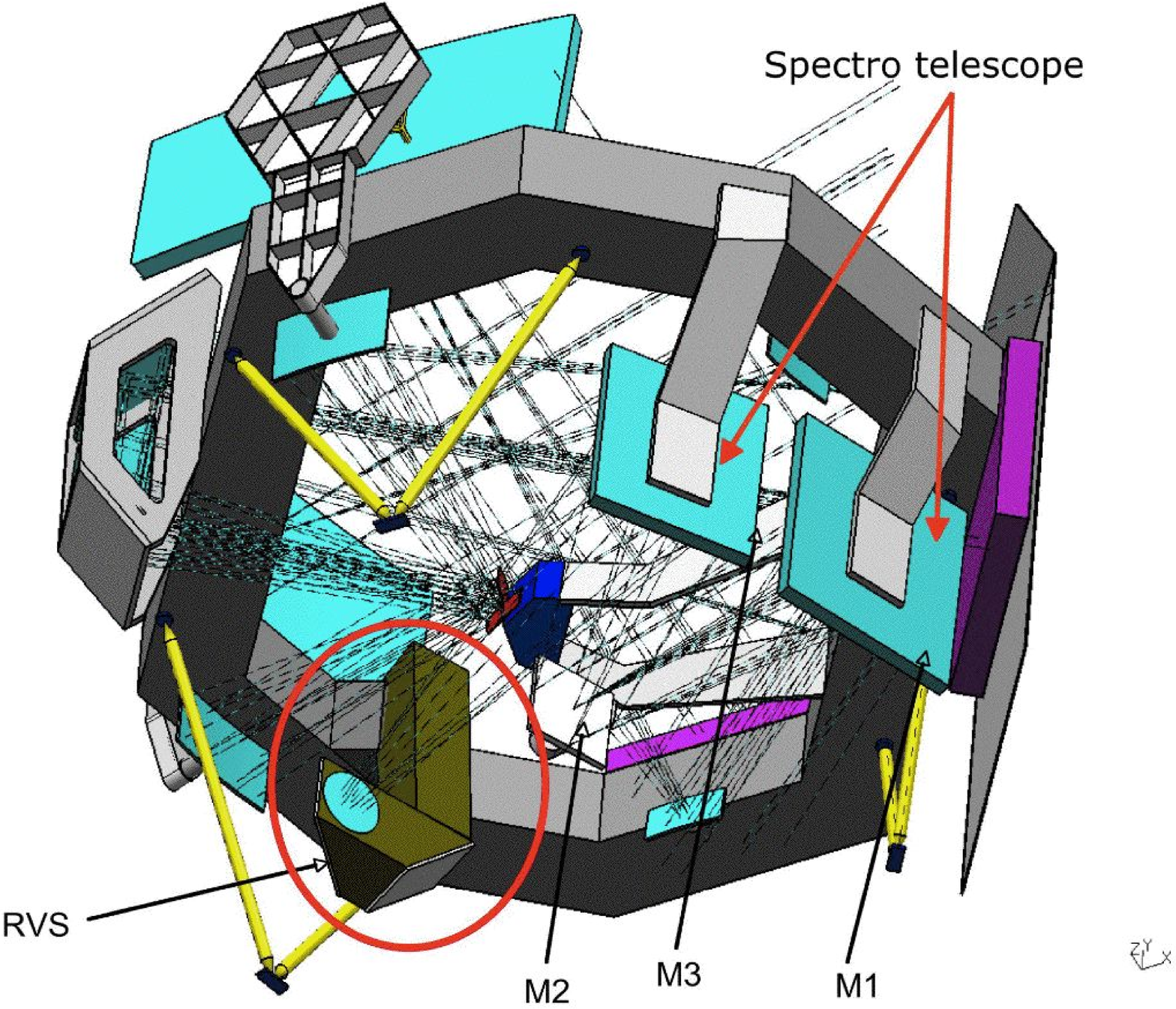}
\includegraphics[height=5.5cm]{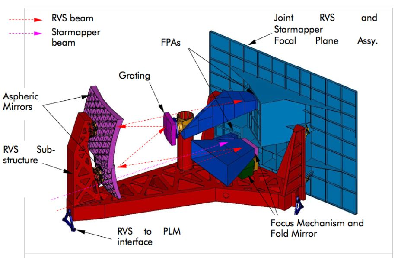}
\includegraphics[height=5.5cm]{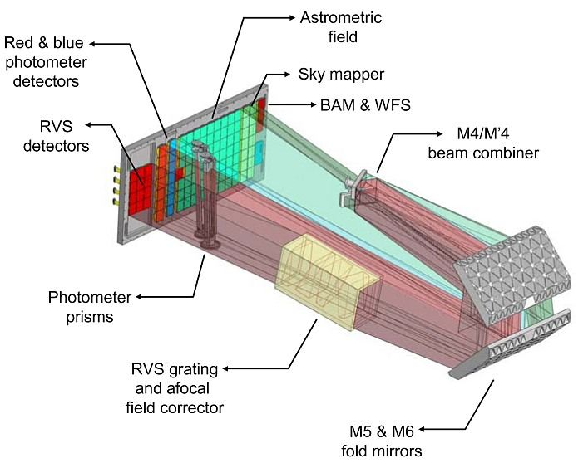}
 }
  \caption{(top) The {\it Gaia}-1 concept. The spectro telescope feeds the RVS assembly circled in red. (middle) The {\it Gaia}-2 design. In this case the optical feed is from the bottom left  onto the fold mirror, and the focal plane with detectors is embedded in the blue radiator.  (bottom) The {\it Gaia}-3 design. Here the telescope feeds the beam combiner M4, before passing through the RVS optics block and on to the 12 RVS detectors on the focal plane. The diagrams at the top and bottom are courtesy of EADS Astrium.}
  \label{fig1}
\end{figure}

Industrial studies for {\it Gaia} commenced in 1997 with a spectrometer incorporated as part of the reference payload. Up until 1999, studies were carried out by Matra Marconi Space (now EADS Astrium) for {\it Gaia}, including the Radial Velocity Spectrometer (RVS), reported in the GAIA CTS Study Final Report GAIA/MMS/TN/037.97. From mid-2001 until mid-2006, spectrometer development was taken forward largely by the RVS Working Group led by D. Katz and U. Munari\footnote{with institutes throughout Europe and in particular in Observatoire de Paris at Meudon, Asiago Observatory, University of Ljubljana,  Mullard Space Science Laboratory of University College London, Cambridge University, Observatoire de la C\^{o}te d'Azur in Nice, Observatoire de Bordeaux, Royal Observatory Belgium, Max Planck Institute for Astrophysics in Garching.} and the RVS Consortium, led by M. Cropper\footnote{at Mullard Space Science Laboratory of University College London, with collaborating institutes GEPI Observatoire de Paris at Meudon (D. Katz), Asiago Observatory (U. Munari) and the University of Ljubljana (T. Zwitter).}. Funding was from national agencies, and two ESA contracts. Development took place in conjunction with both competing prime contractors, EADS Astrium Toulouse and Alcatel Space Cannes, leading to two slightly different designs. Since 2006, the {\it Gaia} payload development including the RVS has been under the auspices of EADS Astrium. These three phases correspond to the {\it Gaia}-1, -2 and -3 designs. They are described in Section 3 below, and diagramatically in Figure~ \ref{fig1}.

\section{Requirements for a {\it Gaia} Spectrograph}

Although the primary driver  for the RVS was initially to furnish radial velocities, the inclusion of a spectrometer adds the possibility for {\it Gaia} to measure astrophysical quantities associated with each star: surface temperature, surface pressure (and hence whether it is a dwarf or a giant star) and metallicity. This adds enormously to the potential science return from {\it Gaia}, especially when allied to the {\it Gaia} photometric data, obtained from the photometer included in the payload, which allows luminosities to be determined. The requirements for velocity and astrophysical measurements are similar, but not completely compatible, and careful assignment of scientific priorities is necessary to ensure maximal return from the instrument. The main parameters to trade off are the spectral resolving power, the spectral range and the radial velocity accuracy. 

Because the spectrometer has to operate within a payload optimised for astrometry, and in particular in scanning mode, this constrains the exposure time. The repeated scanning of the entire sky forces the final spectrum to be combined from the many individual transits. The velocities are obtained by cross-correlation of the spectrum with a template, effectively folding all of the information in the spectrum into a single number: this allows spectra with surprisingly low S/N ratio to yield adequate radial velocities. Because the readout noise from the CCDs (which occurs as each exposure is read out) is the major noise source, exceeding by some margin the cosmic background, sharp attention to minimising it was one of the strongest levers available to maximise the RVS performance. 

While the exposure time can be increased by increasing the spectrometer field of view, the fraction of overlapping spectra will increase rapidly if the spectral resolving power and spectral range is maintained (the scanning prevents the use of a slit in the instrument). Reduction of the spectral range reduces the number of spectral lines available both to anchor the velocity cross-correlation and to provide astrophysical diagnostics, but reduces the overlaps and the cosmic background flux. Reducing the spectral resolving power reduces the crowding and increases the signal per pixel, but reduces the astrophysical parameter diagnostic power, and ultimately as the resolving power is reduced further, the loss of contrast in the lines becomes more important than the increase in signal per pixel. And in order to increase the field of view without increasing the size of the spectrometer focal plane, a smaller image scale is necessary, constraining the nature of the optical feed to the spectrometer, in particular requiring a shorter focal ratio.

\subsection{The Spectral Resolving Power}

The optimal spectral resolving power for the RVS received substantial attention by the RVS Working Group. Resolving powers below $R\sim5000$ would impact the astrophysical parameter determination disproportionately, while those above $R\sim20000$ were clearly sub-optimal for the radial velocities, and would limit the faintest objects amenable for astrophysical parameter determination to only a bright subset. Simulations were carried out for different spectral types, including the effect of spectral overlaps in crowded fields. These indicated that the overlaps became a limiting factor in regions of moderate and greater crowding for higher resolutions. Finally,  $R\sim10000$ was adopted as  an optimal compromise, sufficiently high to determine the $[\alpha/{\rm Fe}]$ ratio in addition to the overall metallicity, while retaining almost optimal radial velocity performance at the faint end.

\subsection{The Spectral Range}

The initial Matra Marconi studies took as reference the spectral range around the MgII complex at 500-550 nm, as suggested by Favata \& Perryman (1995), but in 1997, Munari suggested that the wavelength interval be changed to the interval around the Ca II triplet near 850 nm. His argument was that this region held the strongest lines in the red/near IR part of the spectrum, it was not affected by telluric absorption (in the atmosphere) so that associated ground-based studies could be carried out, there was little contamination by interstellar components, and most importantly, that as well as the strong Ca lines, it was a region rich in line species including FeI, TiI, MgI, SiI and Paschen lines (important for hotter stars). These arguments proved convincing, and this region was adopted at that time for RVS. 

The spectral range clearly needed to incorporate all three Ca lines, but not much more, to minimise overlaps and background. It was set to a range $847-874$ nm  by the Working Group to include important Fe lines, and to provide sufficient continuum outside of the Ca triplet.

\subsection{Radial Velocity Accuracy and Limiting Magnitude}

The initial radial velocity performance identified in Favata \& Perryman (1997) was put on a firmer footing by the RVS Working Group, in continuing studies (eg Munari {\it et al.} 2002,  Zwitter, 2002, Wilkinson, 2002) and then most comprehensively in two papers Katz {\it et al.} (2004) and Wilkinson {\it et al.} (2005). The radial velocity accuracy requirements were set by requiring them to be equivalent to the tangential accuracies from astrometry for giant star (mostly KIII) tracers at large distances, and the limiting magnitude was similarly set by requiring these to be observable across the Galaxy. These requirements were driven mostly by halo and thick disk science and, in summary, these required accuracies of $<15$ km s$^{-1}$ for $V=17-18$, with tighter accuracies for brighter magnitudes, typically $<3$ km s$^{-1}$ for $V=14$. This was revised to constrain brighter stars, to  $<1$ km s$^{-1}$ for $V=13$ in the Mission Requirements Specification during early 2005.

\section{{\it Gaia} Design Evolution}

\vspace*{-2mm}
\subsection{{\it Gaia}-1}

The {\it Gaia}-1 design in 1999 resulting from the Matra-Marconi CTS study was a lens-based (dioptric) system, with a collimator, grating and camera. The  instrument was fed by a dedicated {\sc spectro} telescope of $0.7\times0.75$ m$^2$ aperture, with a large ($1^{\circ}$ square) field of view and moderate f/ratio, so that the spectra scanned slowly over the focal plane, maximising the exposure time. The dispersion direction was set perpendicular to the scan direction, which caused some difficulties, as the periodic across-scan motion from the forced precession of the satellite then compromised the spectral resolution. There were six narrow red-sensitive CCDs in the RVS focal plane, with $10\times15 \mu$m pixels, especially developed for {\it Gaia}.

\vspace*{-2mm}
\subsection{{\it Gaia}-2}

The {\it Gaia}-2 design made a number of enhancements. An important decision was to place the dispersion direction in the scan direction (Cropper \& Mason, 2001). A range of dioptric designs was explored, to minimise distortion, mass, space envelope and manufacturing complexity while maintaining the image quality. Ultimately the optical design evolved to a pared-down reflective design consisting of only concave aspheric collimator and camera mirrors, and a convex spherical diffraction grating in an Offner-relay, which maximised throughput and image quality. This was optimised separately to the Alcatel Space and EADS Astrium  payloads

The focal plane was also an area where substantial tradeoff studies were carried out.  Several different options were considered, depending on the number of CCDs in the along-scan direction,  whether L3CCDs or conventional CCDs were to be used, whether a mechanism was to be incorporated to compensate for the scan law across-scan motion and whether there would be on-board processing to combine the data from different CCDs. The outcome of this tradeoff was that configurations with more narrow CCDs of the L3CCD type were preferred. L3CCDs are CCDs with on-chip amplification of the signal before the readout node: this effectively reduces the readout noise to negligible levels, considerably enhancing the RVS performance. A space qualification of these devices was carried out to ensure their suitability (Smith, Ingley \& Holland 2005; Cropper et al. 2005b).

In order to comply with data downlink budgets, the {\it Gaia}-2 RVS combined the data from individual CCDs on board, and both hardware and software schemes were developed to address this. Regions of the sky not wanted were simply cut from the combined image, but with the 2-D relationship of all spectra maintained.

The final configuration of this design is in Cropper et al (2005a).

\vspace*{-2mm}
\subsection{{\it Gaia}-3}

In 2005/6, a new configuration was proposed by the winning EADS Astrium consortium. This RVS uses the main astrometric telescopes, windowing of the spectra at CCD level as well as binning in the along-scan direction by 3 pixels for fainter stars, reducing their spectral resolution. All but the brightest spectra are collapsed to 1-D at the time of readout. There is no on-board processing. Non-L3CCDs are used (red sensitive variants of the astrometric devices). The advantages of this configuration (from the RVS perspective) is that there is less crowding of images and no on-board processing is required. On the other hand, two spectral resolutions (HR/LR) are required to minimise the readout noise; there are difficulties in dealing with spectral overlaps using the 1-D data; exposures are short (from the large image scale) and non-L3CCDs lead to reduced performance (~1 mag); the radiation damage effects are significantly worse because of wide CCDs, and fewer transits are recorded. 

Radiation damage in the CCDs causes the leading edge of spectrum to be consumed by traps in radiation-damaged Si within pixels. This leads to a magnitude-dependent radial velocity bias, to changes to the equivalent width of spectral lines, reduced S/N ratio from charge loss. However, tests have shown that even very low signals (1 e$^{Ð}$) are recovered after 4500 line shifts in the CCD. The damage effects can be modelled using fast phenomenological models (eg CDM02, Short 2009)  and post-calibrated. Besides these previously expected effects, two new technical challenges have arisen. As a result of the on-chip windowing and binning,  the stability of the CCD readout speed is lost, leading to a PEM-CCD bias instability which shifts the baseline of the spectra. The second is a serial register charge transfer inefficiency. Each will be addressed by combination of calibration and software measures.

\section{Summary}

We have described how in the {\it Hipparcos} programme the absence of the radial velocities in an astrometric mission was considered to limit the science return and that the  RVS was included in the {\it Gaia} instrument suite to address this shortcoming. Its inclusion endows {\it Gaia} with a complete suite of instruments to measure kinematic and astrophysical properties of stars in the Galaxy.

Given that the photons from spectra will be distributed over a much larger number of pixels than in the imaging data, the photon-starved nature of the instrument (low exposure levels) is an unusual parameter space to be optimised for a scientific instrument. This has given rise to many unique challenges. In particular, very careful attention has had to be paid to noise sources. Further, the extended length of the spectra require strategies for minimising the data rate; in addition, because of the longer data windows, source overlapping in crowded regions is more common and problematic. 

Over the conception and development of {\it Gaia}-RVS many different configurations and technologies have been examined, tested and built to address the science requirements. The process has been subject to many considerations: technical, programmatic, sociological in nature. Contributions over many years have been made by science groups in academia, by industry and by ESA: a massive effort that is a tribute to the dedication and organisation across Europe in these interdependent spheres.

In the process of development and in the face of real-world constraints, some capabilities have been lost, but in the end, {\it Gaia}-RVS will broadly do what was set down for it more than a decade ago: providing a massive resource of $\sim10^8$ multi-epoch spectra as part of the {\it Gaia} scientific bounty.

\section*{Acknowledgements}
We are grateful to Catherine Turon for the invitation to write this manuscript at the ELSA workshop. The historical sections have drawn on the private communications from Catherine Turon and Michael Perryman. MC thanks the organisers for financial support.


\end{document}